\documentclass{ws-procs9x6}            

\usepackage{graphicx}

\def\bra{\langle}
\def\ket{\rangle}

\def\half#1{{#1\over 2}}

\def\crsth{{[8\half3]$_8$}}
\def\crsh{{[8\half1]$_8$}}
\def\cwsh{{[8\half1]$_1$}}
\def\crshf{{[1\half1]$_8$}}

\def\Lamc{\mbox{$\Lambda_c$}}

\def\Sigcstar{\mbox{$\Sigma_c^*$}}

\def\Jpsi{\mbox{$J\!/\!\psi$}}
\def\etac{\mbox{$\eta_c$}}
\def\Dbar{\mbox{$\overline{D}$}}
\def\Dbarstar{\mbox{$\overline{D}{}^*$}}

\def\Qbar{\overline{{Q}}}

\def\half#1{\text{${#1\over 2}$}}

\def\Vcoul{V_\text{Coul}}
\def\Vconf{V_\text{conf}}
\def\Vcmi{V_\text{CS}}

\begin{document}
\title{Strange pentaquarks with a hidden heavy quark-antiquark pair}

\author{Sachiko Takeuchi$^{1,2,3}$, Alessandro Giachino$^{4,5}$, Makoto Takizawa$^{6,2,7}$,\\ 
Elena Santopinto$^{4}$ and Makoto Oka$^{8}$}

\address{$^1$Japan College of Social Work, Kiyose, Tokyo 204-8555, Japan\\
E-mail: s.takeuchi@jcsw.ac.jp
}
\address{%
$^2$Theoretical Research Division, Nishina Center, RIKEN,\\
 Hirosawa, Wako, Saitama 351-0198, Japan
}
\address{%
$^3$Research Center for Nuclear Physics (RCNP), Osaka
University,\\
 Ibaraki, Osaka 567-0047, Japan
}
\address{%
$^4$Istituto Nazionale di Fisica Nucleare (INFN), \\
Sezione di
Genova, via Dodecaneso 33, 16146 Genova, Italy
}
\address{%
$^5$Dipartimento di Fisica dell'Universit\`a di Genova, \\
via Dodecaneso 33, 16146 Genova, Italy
}
\address{%
$^6$Showa Pharmaceutical University, \\
Machida, Tokyo
194-8543, Japan
}
\address{%
$^7$J-PARC Branch, KEK Theory Center, \\
Institute for Particle and Nuclear Studies, KEK, 
Tokai, Ibaraki 319-1106, Japan
}
\address{%
$^8$Advanced Science Research Center, Japan Atomic Energy Agency, \\
Tokai, Ibaraki 319-1195, Japan}

\begin{abstract}
The strange pentaquarks with hidden heavy quark pair ($q^3c\bar c$ and $q^3b \bar b$)
are investigated by the coupled-channel quark cluster model.
Two types of the $q^3$ color-octet configurations
are found to provide the attraction,
which makes bound states, sharp resonances, and cusps in the baryon meson scattering.
A resonance appears at around 4500 MeV in the strange hidden charm sector.
Such structures are more clearly seen in the hidden bottom systems.
\end{abstract}

\bodymatter

\section{Introduction}\label{aba:sec1}
Recently,
the $P^+_c(4312) $ state and the two narrow peaks, $P^+_c (4440)$ and $P^+_c(4457)$, 
have been found in 
$\Lambda^0_b\rightarrow J/\psi K^-p$ decay 
by LHCb \cite{Aaij:2015tga,Aaij:2019vzc}.
Many theoretical works have been proposed to investigate the hidden charm systems 
 ({\it e.g.}\ see ref\cite{Chen:2016qju}).
In ref.\cite{Takeuchi:2016ejt}, it has been reported that the quark cluster model 
gives a bound state, resonances, and a cusp 
in the $N\Jpsi$ or $Y_c {\bar D}$ scattering. 
Together with the pion exchange, 
the model seems to reproduce the observed peaks \cite{Yamaguchi:2017zmn,Yamaguchi:2019seo}.
Those structures appear because of 
the three light quarks in the color-octet isospin-1/2 spin-3/2 compact configuration.
Such a configuration appears when the baryon ($q^2Q$) and the meson ($q\bar Q$) come close to each other,
and produces a channel-dependent short-range interaction between the hadrons.

Here we discuss the strange pentaquarks with the hidden heavy quark pair, $udsQ\bar Q$. 
Now the three light quarks 
can also be color-octet and flavor-singlet. 
Note that this flavor-singlet configuration for the $S$-wave three quarks 
can form only when the quarks are color-octet.
In this work we show that this configuration also
causes resonances. They are more clearly seen in the hidden bottom sector,
whereas the one in the charm sector may be observed, {\it e.g.},  in $\Lambda^0_b\to J/\psi \Lambda \phi$ decay
or in $\Xi_b^- \to J/\psi \Lambda K^-$ decay.

\section{Method}

Let us first classify possible configurations of the three light quarks in a $q^3 Q\bar Q$ system.
Since the whole system is color-singlet and the hidden heavy quark part ($Q\bar Q$) is color-singlet or octet,
the remaining light quarks ($q^3$) are also color-singlet or octet.
A color-singlet $q^3$ belongs to the usual flavor spin {\bf 56}$_{f\sigma}$ multiplet, 
which consists of the flavor-octet 
and the flavor-decuplet 
baryons.
A color-octet $q^3$ 
belongs to the {\bf 70}$_{f\sigma}$ multiplet,
which consists of 
flavor-singlet spin-1/2,
the flavor-octet spin-1/2,
the flavor-octet spin-3/2,
and flavor-decuplet spin-1/2 configurations\cite{Irie:2017qai}:
\begin{align}
{\bf 56}_{f\sigma}&={\bf 8}_f\times{\bf 2}_{\sigma}+{\bf 10}_f\times{\bf 4}_{\sigma}
\label{eq:56}
\\
{\bf 70}_{f\sigma}&={\bf 1}_f\times{\bf 2}_{\sigma}+{\bf 8}_f\times{\bf 2}_{\sigma}+{\bf 8}_f\times{\bf 4}_{\sigma}+{\bf 10}_f\times{\bf 2}_{\sigma}~.
\label{eq:70}
\end{align}
The color spin interaction for the quarks contributes differently to the two terms in rhs of eq.\ (\ref{eq:56}),
which causes the hyperfine splitting (HFS) of the baryon masses.
It also contributes to the terms in eq.\ (\ref{eq:70}), which gives the HFS of the pentaquarks, $q^3Q\bar Q$,
at the heavy quark limit (CS):
\begin{align}
{\rm CS}&= -\bra q^3[fs]_c| 
\sum
(\lambda\cdot\lambda)(\sigma\cdot\sigma)
|q^3[fs]_c\ket ,
\label{eq:cmi}
\end{align}%
where $q^3[fs]_c$ stands for the $q^3$ configuration with the flavor $f$, spin $s$, and color $c$. 
The difference between CS and the one that contributes to the threshold energy (CS$_T$)
shows whether the color-spin term
of the Hamiltonian works as an attractive force (CS$-$CS$_T<0$) or not ($>0$).
For the isospin-0 $udsQ\bar Q$,
the terms with ${\bf 1}_f$ or ${\bf 8}_f$
are relevant.
In Table \ref{tbl:cmi}, for each $[fs]_c$ and
the spin of $Q\bar Q$, $S_{Q\bar Q}$, 
we list possible total spin, $J$, CS,
the relevant lowest $S$-wave threshold with CS$_T$.
\begin{table}[t]
\renewcommand\arraystretch{1.1}
\tbl{The classification of the isospin $\half1$ strangeness 0
and isospin 0 strangeness $-1$ $q^3{Q\bar Q}$ states.
$[fs]_c$ stands for $q^3$ with  flavor, $f$, spin $s$, and  color $c$.
$S_{Q\bar Q}$ stands for the ${Q\bar Q}$ spin, and  
$J$ for the total spin of  $uud{Q\bar Q}$.
The lowest $S$-wave threshold for $Q=c$, CS, and CS$_T$  are also listed.}
{\begin{tabular}[\textwidth]{ccccccccccccccccccccccc}\toprule
[$fs]_c$&  CS  &  $S_{Q\bar Q}$ & $J$ & \multicolumn{2}{c}{the lowest $S$-wave threshold} & CS$_T$ & CS$-$CS$_T$
\\\colrule
\cwsh\   &$-8$& 0 & $\half1$ &$N\etac$ & $\Lambda\etac$ & $-8$ & 0
\\
&&  1 & $\half1$, $\half3$ &$N\Jpsi$& $\Lambda\Jpsi$&
\\\colrule
\crshf\ & $-14$& 0 & $\half1$ & &  $\Lambda_c D_s$& ${-8}$& $-6$
\\
&&  1 & $\half1$, $\half3$ & & $\Lambda_c D_s{}^{(*)}$&
\\\colrule
\crsh\ & $-2$& 0 & $\half1$ & \Lamc\Dbar&\Lamc$D_s$&$-8$& $6$
\\
&& 1 & $\half1$, $\half3$ & \Lamc\Dbar${}^{(*)}$ &\Lamc $D_s{}^{(*)}$&
\\\colrule
\crsth\  &\phantom{$-$}2& 0 & $\half3$ &   \Sigcstar\Dbar &$\Xi_c' \bar D$ & ${8\over 3}$& $-{2\over 3}$
\\
&&  1 & $\half1$,$\half3$,$\half5$  &   $\Sigma_c{}^{(*)}$\Dbar${}^{(*)}$& $\Xi_c' \bar D{}^{(*)}$, $\Xi_c^* {\bar D}{}^*$ &
\\\botrule
\end{tabular}}
\label{tbl:cmi}
\end{table}%
It is found that the configurations of 
$[fs]_c$=$[1\half1]_8$ and $[8\half3]_8$ 
provide attraction. 

In order to understand if such an attraction is strong enough 
to produce resonances or bound states,
we investigate the situation dynamically by employing 
the coupled-channel quark cluster model
for the $uudQ\Qbar$ $I(J^P)=\half1(\half1^-,\half3^-,\half5^-)$ 
and
the $udsQ\Qbar$ $I(J^P)=0(\half1^-,\half3^-,\half5^-)$ 
systems with $Q=c$ and $b$.
The model is essentially the same as used in 
ref.\ \cite{Takeuchi:2016ejt} with the difference 
that now we are  considering also the strange and bottom quarks. 
The Hamiltonian for quarks, 
$H_q$, consists of the kinetic term, $K$, the confinement term, $\Vconf$, the 
color Coulomb term, $\Vcoul$, 
and the color spin term, $\Vcmi$:
\begin{align}
H_q&=K+\Vconf+\Vcoul+\Vcmi\ .
\end{align}
Both of $\Vcoul$ and $\Vcmi$ are considered 
to come from the effective one-gluon exchange interaction
between quarks.
%
The $s$ and $b$ quark masses used in our model 
and the obtained 
 hadron masses are listed in Table \ref{tbl:hadron_masses}.
The hadron masses are reproduced  well,
but to use the observed threshold energies, we introduce the mass correction to the kinetic term.
In the long range region,
this model becomes 
essentially 
a free baryon-meson model.
In the short range region,
 an interaction between the baryon and the meson 
arises from the quark degrees of freedom and from the interaction between quarks.

\begin{table}[t]
\renewcommand\arraystretch{0.95}
\tbl{%
The quark masses used in the model and the obtained hadron masses (in MeV). 
The numbers in parentheses are the isospin-averaged experimentally observed masses\cite{Tanabashi:2018oca}.
}
{\begin{tabular}{cccccccccccccc}\toprule
$m_u(=m_d)$ &$m_s$ &$m_c$ &$m_b$ \\
300 & 510 &1741.5 &5110.9\\ \colrule
%
%
N              &
$\Lambda$      &
$\Sigma$       &
$\Sigma^*$     &
$\Xi$          &
$\Xi^*$        &
\\
~922(~939)& 
1110(1116)& 
1188(1193)& 
1394(1385)& 
1328(1318)& 
1523(1533)& 
\\\colrule
$\Lambda_c$    &
$\Sigma_c$     &
$\Sigma_c^*$   &
$\Xi_c$        &
$\Xi_c'$       &
$\Xi_c^*$      \\
2292(2286)& 
2454(2454)& 
2516(2518)&
2495(2469)&
2578(2579)&
2630(2646)\\
\colrule 
$\Lambda_b$    &
$\Sigma_b$     &
$\Sigma_b^*$   &
$\Xi_b$        &
$\Xi_b'$       &
$\Xi_b^*$      \\
5640(5620)& 
5826(5813)& 
5847(5833)& 
5827(5794)&
5936(5935)&
5952(5954)
\\ \colrule 
$\eta_c$       &
\Jpsi\         &
\Dbar\         &
\Dbarstar\     &
$D_s$     &
$D_s^*$     \\
2981(2984)&
3101(3097)&
1863(1867)&
2005(2009)&
1969(1968)&
2112(2112)
\\
\colrule
$\eta_b$       &
$\Upsilon$         &
$B$        &
$B^*$     &
$B_s$     &
$B_s^*$     \\
9395(9399)&
9460(9460)&
5276(5279)&
5331(5325)&
5356(5367)&
5417(5415)
\\
\botrule
\end{tabular}}
\label{tbl:hadron_masses}
\end{table}

\begin{table}[t]
\renewcommand\arraystretch{1.15}
\tbl{The number of structures in the baryon meson scattering with the number of the 
relevant $q^3$ configurations which provide the attraction.}
{\begin{tabular}{cccccccccccccc}\toprule
 $S$ & $J$  & [8\half3]$_8$ & $uudb\bar b$ & $uudc\bar c$ \\ \colrule
0& \half1   & 1 & 1 & 1 \\
 & \half3   & 2 & 2 & 1  \\
 & \half5   & 1 & 1 & 1 
\\ \botrule
\end{tabular}~
\begin{tabular}{cccccccccccccc}\toprule
 $S$ & $J$ &  [1\half1]$_8$& [8\half3]$_8$ & $udsb\bar b$ & $udsc\bar c$\\ \colrule 
$-1$& \half1  & 2 & 1 & 3 & 1\\
    & \half3  & 1 & 2 & 3 & 1 \\
    & \half5  & 0 & 1 & 1 & 1
\\\botrule
\end{tabular}
}
\label{tbl:results}
\end{table}%

\section{Results}
In Table \ref{tbl:results},
we summarize 
the number of the bound states and resonances
that appear  in
the quark cluster model calculation.
Resonances we count here are the ones that
the diagonal phase shift rises more than $\pi/2$.
The number of the bound states and resonances roughly corresponds to the 
number of quark configurations that provide the attraction.

Correspondence between the attractive
color-octet configurations and the structures in the scattering 
is more prominent in the $q^3b\bar b$ systems.
As shown in  Fig.\ \ref{fig:bbbar}, for example,
the $uudb\bar b$ $J=3/2$ system has two resonances, which corresponds to the
two configurations of [8\half3]$_8$.
In $udsb\bar b$ $J=3/2$ system has one more resonance at the $\Lambda_b B_s^*$ threshold,
which corresponds to [1\half1]$_8$.
There are one bound state in each of the $q^3Q\bar Q$ $J=5/2$ systems,
where the three light quark configuration is
[8\half3]$_8$.

The resonances can still be seen in the $q^3c\bar c$ systems.
As for the spin-3/2 $udsc\bar c$ systems,
there is a resonance at around 4500 MeV,
in which both of the [1\half1]$_8$ and [8\half3]$_8$ configurations
contribute (Fig.\ \ref{fig:ccbar}).
Energetically, it can be seen 
in $\Lambda^0_b\to J/\psi \Lambda \phi$ decay
or in $\Xi_b^- \to J/\psi \Lambda K^-$ decay.

The energies and channels in which the resonances appear may vary by the choice of the potential parameters.
The existence of the peaks, however, seems to be robust. 
Investigating the pentaquark systems will show us properties of the color non-singlet configuration,
and will give us information to understand features of the low energy non perturbative QCD.

{\small This work is supported in part by 
by JSPS KAKENHI No.\ 16K05361.}
\vspace*{-5mm}

\newlength{\figw}
\setlength{\figw}{0.48\textwidth}

\begin{figure}[t]
\begin{center}
\includegraphics[width=\figw,bb=0 0 515 273]{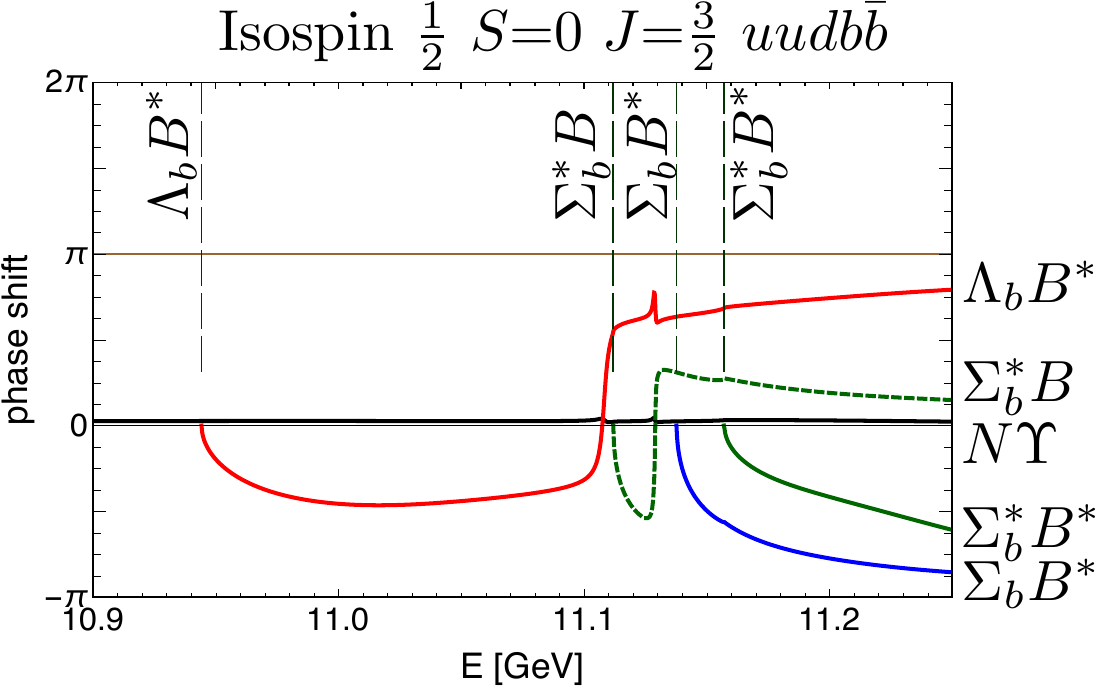}\hfill
\includegraphics[width=\figw,bb=0 0 515 273]{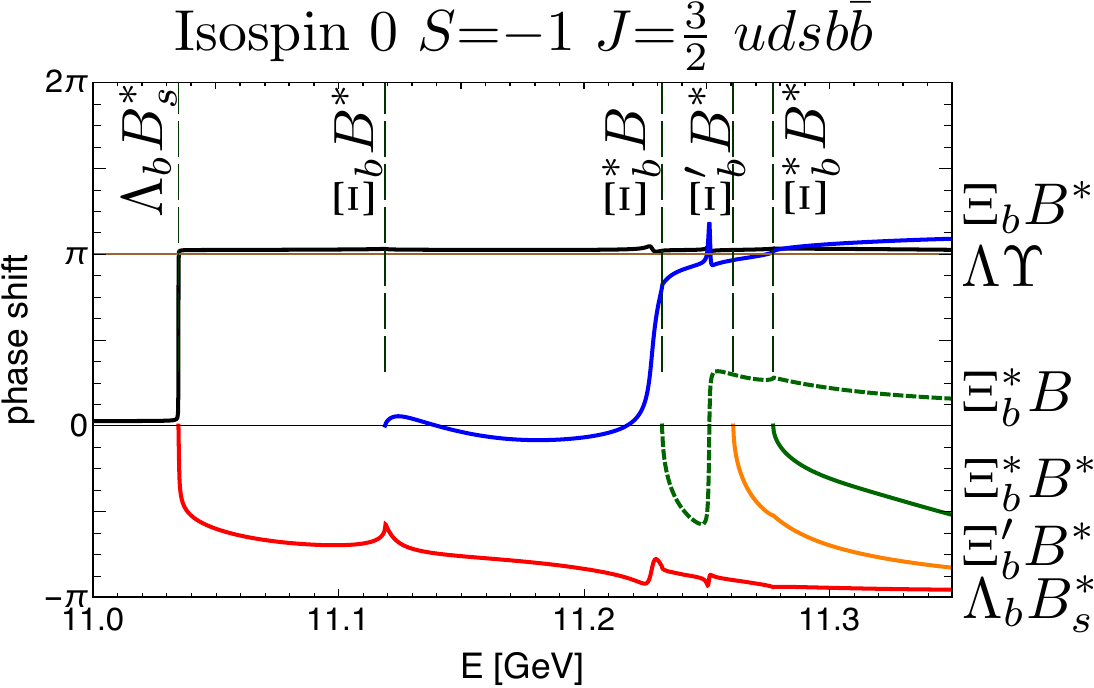}\\\vspace{-3mm}
\end{center}
\caption{Hidden bottom baryon meson scattering phase shifts of the isospin 1/2 (left) 
and the strangeness $-1$ isospin 0 systems (right). Total angular momentum is 3/2.}
\label{fig:bbbar}
\end{figure}

\begin{figure}[t]
\begin{center}\vspace{2mm}
\includegraphics[width=\figw,bb=0 0 515 273]{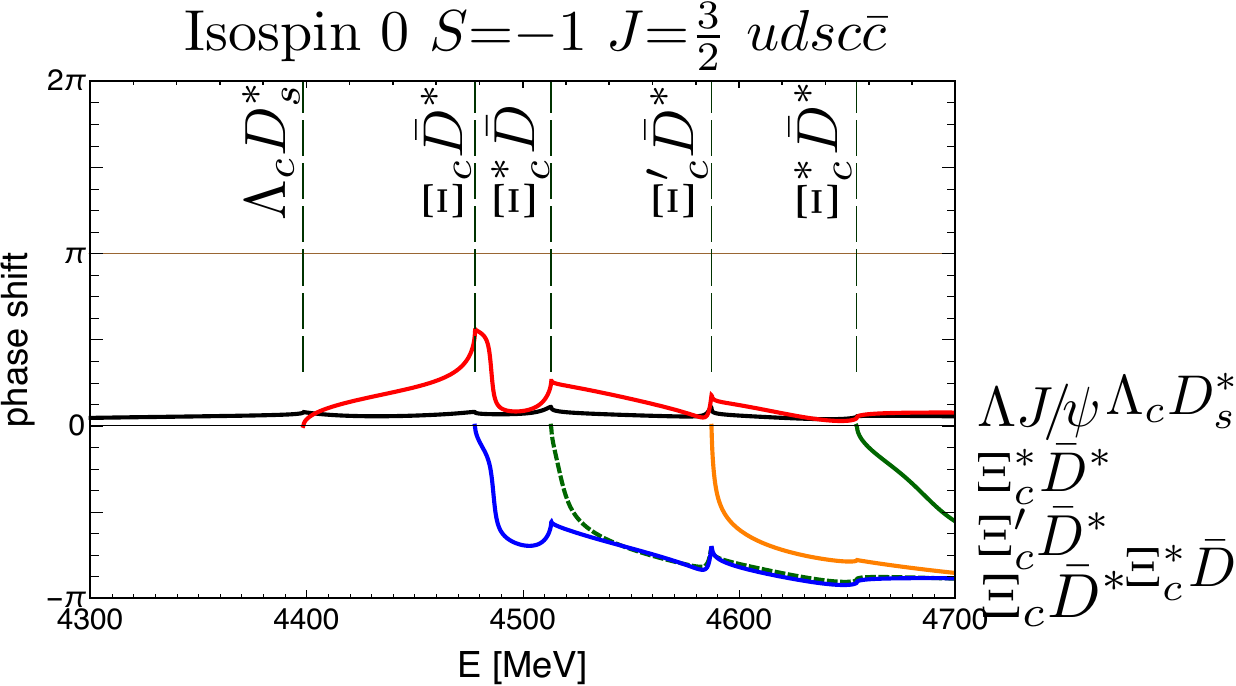}\\\vspace{-3mm}
\end{center}
\caption{Hidden charm baryon meson scattering phase shifts of the strangeness $-1$ isospin 0 systems. Total angular momentum is 3/2.}
\label{fig:ccbar}
\end{figure}

{
\itemsep=0mm

}

\end{document}